\documentclass{ifacconf}
\usepackage{natbib}        
\usepackage{xcolor}
\usepackage{amsmath} 
\usepackage{amssymb}
\usepackage{graphicx,kantlipsum,setspace}
\usepackage{caption}
\usepackage[font=scriptsize]{caption}\usepackage[font=scriptsize]{caption}

\newcommand{\nd}[1]{\mathsf{#1}}

\begin{document}
\begin{frontmatter}

\title{Excitable crawling\thanksref{footnoteinfo}} 

\thanks[footnoteinfo]{This work was supported by the Schmidt Science Postdoctoral Fellowship from Schmidt Futures and the Rhodes Trust.}

\author[First]{Juncal Arbelaiz} 
\author[Second]{Alessio Franci} 
\author[First]{Naomi Ehrich Leonard}
\author[Third]{Rodolphe Sepulchre}
\author[Fourth]{Bassam Bamieh}

\address[First]{Mechanical and Aerospace Engineering Department, Princeton University, Princeton, NJ 08544 USA (e-mail: jarbelaiz@schmidtsciencefellows.org)}
\address[Second]{Department of Electrical Engineering and Computer Science, University of Liège, B-4000 Liège, Belgium, and WEL Research Institute, Wavre, Belgium.}
\address[Third]{Department of Electrical Engineering, 
   KU Leuven, B-3001 Leuven, Belgium}
   \address[Fourth]{Mechanical Engineering Department, 
   UCSB, Santa Barbara, CA 93106-5070 USA}

\begin{abstract}                
We propose and analyze the suitability of a spiking controller to engineer the locomotion of a soft robotic crawler. Inspired by the FitzHugh-Nagumo model of neural excitability, we design a bistable controller 
with an electrical flipflop circuit representation 
capable of generating spikes on-demand when coupled to the passive crawler mechanics.
A proprioceptive sensory signal from the crawler mechanics 
turns bistability of the controller into a rhythmic spiking.
The output voltage, in turn, activates the crawler’s actuators to generate movement through peristaltic waves. 
We show through geometric analysis that this control strategy achieves endogenous crawling. The electro-mechanical sensorimotor interconnection
provides embodied negative feedback regulation, facilitating 
minimal controller tuning for 
locomotion.
Dimensional analysis  provides insights on the characteristic scales in the crawler’s mechanical and electrical dynamics, and how they determine the crawling gait. Adaptive control of the electrical scales to optimally match the mechanical scales can be envisioned to achieve further efficiency, 
as in 
homeostatic regulation of 
neuronal circuits.
Our approach can 
scale up to multiple sensorimotor loops inspired by
biological central pattern generators.
\end{abstract}


\end{frontmatter}

\section{Introduction}
\vspace{-.2cm}
In nature, the elasticity of the musculoskeletal system is a fundamental ingredient for the effectiveness of animal performance in a wide range of tasks, \cite{DellaSantina2023}. Inspired by this observation, in engineering,  soft robots aim to leverage the compliance of their bodies to efficiently handle different environments and uncertainty. While their promise is large, the variability in robot morphology, the complexity of creating accurate models of their dynamics, and the limited onboard power make the design and implementation of feedback control architectures challenging. Consequently, soft robotics is a technology still in its infancy. 

Among the different strategies that soft-bodied animals and robots utilize for locomotion, crawling is a common one.  Crawlers typically move by propagating peristaltic waves along their bodies, with alternating contraction and relaxation of muscles or actuators -- see, e.g., \cite{Paoletti2014}. In this work, we evaluate the use of an 
excitable feedback controller -- \cite{Sepulchre2018}, \cite{Sepulchre2019}, \cite{Sepulchre2022}  -- in a soft one-segmented robotic crawler to engineer its peristaltic locomotion.
The resulting {\it spiking}  controller is inspired by the FitzHugh-Nagumo model of neural excitability.

Spiking control systems combine the best properties of analog and digital controllers: they offer robustness to model uncertainty and component degradation, they can be modulated for  control across scales, and they possess the continuous adaptation capability of analog systems with the discrete reliability of digital automata; all with reduced power consumption -- see \cite{Sepulchre2022}. These features are very well suited to the inherent challenges of synthesis and implementation of feedback control for soft robotics, making spiking control a promising alternative to current approaches.

This work analyzes a single segmented soft robotic crawler, for which a single-input single-output spiking controller is designed. 
The main contribution  is to show, through a blend of dimensional analysis, geometric singular perturbation, and numerical simulations, that a proprioceptively modulated 
bistable feedback controller is able to produce self-sustained closed-loop electromechanical oscillations,
leading to peristaltic locomotion. 
Extensions to multi-segmented crawlers will require the synthesis of multi-input multi-output spiking controllers, which can rely on distributed control theory and the recent theory of fast and flexible multi-agent multi-option decision-making -- \cite{Leonard2024,bizyaeva2022nonlinear}.





\section{Excitable crawler control}

\subsection{Crawler dynamics}
\begin{figure}[h]
    \centering
    \includegraphics[width=8.4cm]{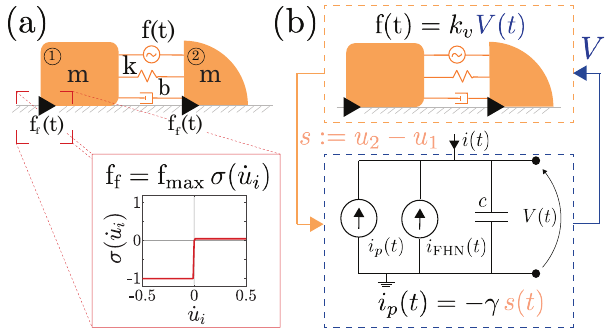}
    \caption{ 
    (a) Schematic of the soft crawler analyzed in this work. The 
    panel represents the 
    nonlinear anisotropic friction model as given in \eqref{eq:sigma_def}. (b) Closed-loop system: information flow between the crawler and the neuromorphic excitable controller proposed in Section \ref{subsec:excitable_controller}. The neuromorphic controller has the
    realization of a flip-flop 
    electrical circuit.}
    \label{fig:crawler_schematics}
\end{figure}

 We analyze a soft crawler, whose body is composed of a single segment of natural length $l_0$. Fig. \ref{fig:crawler_schematics}a) provides a schematic of the crawler. Its dynamics are modeled as
\begin{subequations}
\begin{align}
  m  \ddot u_1 &= k (u_2-u_1 ) + b (\dot u_2 - \dot u_1)  - f_f(\dot u_1 )-f,
  \label{eq:crawler_dynamics_1}\\
  m  \ddot u_2 &= k (u_1-u_2) + b (\dot u_1 - \dot u_2)  - f_f(\dot u_2) + f,
   \label{eq:crawler_dynamics_2}
\end{align}
\label{eq:dimensional_crawler_dynamics}
\end{subequations}
where $u_i$ denotes the displacement of mass $i$ ($i = 1,2$). The time variable is denoted by $t$ and we use $\dot{u}_i$ to denote $\mathrm{d} u_i/\mathrm{d} t$. The viscoelasticity of the crawler's body is captured through the elastic constant $k$ and the viscous damping constant $b$. $f$ is the actuator signal and $f_f$ the frictional force arising due to the interaction of the crawler with its environment. $f_f$ is a nonlinear  anisotropic function of the local speed, modeled as $f_f(\dot{u}) = f_{max} \, \sigma(\dot{u})$, where $f_{max}$ denotes its amplitude and 
\begin{equation}
    \sigma(\dot{u}):= \frac{\tanh{( \dot{u}/\epsilon_f + n_f) - \tanh{(n_f)}}}{1 + \tanh{(n_f)}}.
    \label{eq:sigma_def}
\end{equation}
Note that $\sigma(0) = 0$ and that $\sup_{\dot{u}} |\sigma(\dot{u})| = 1$. The parameters $n_f$ and $\epsilon_f$ tune the anisotropy of the frictional force and its slope, respectively. Friction anisotropy introduces a preferential direction of motion for the crawler: forward motion experiences less friction than backward motion. The frictional force model as a function of the local speed $\dot{u}_i$ is depicted in Fig. \ref{fig:crawler_schematics}a).

\vspace{-.4cm}
\subsection{A neuromorphic excitable controller}
\label{subsec:excitable_controller}
We aim to design $f$ in \textit{proprioceptive feedback} to yield \textit{endogenous crawling} -- that is, crawling motion without 
exogenous periodic actuator commands. To design such a controller, we take inspiration from \textit{neuronal dynamics}.

We let the actuator signal $f$ be proportional to the output voltage $V$ of an electrical circuit 
(Fig. \ref{fig:crawler_schematics}b): $$f = k_v \, V,$$ 
where $k_v \in \mathbb{R}_+$ denotes the voltage gain. We design the electrical circuit to contain two properties whose intertwined action will enable endogenous crawling:\\
     \textit{(1) Bistability}, which is achieved by introducing a
     voltage-controlled current source ($i_{\text{FHN}}$) with a localized region of negative conductance, inspired by the FitzHugh-Nagumo model of neuronal excitability -- see \cite{FitzHugh1955}:
     $$
         i_{\text{FHN}} = - \alpha \, V^3 + \beta \, V,
         \label{eq:i_FHN}
     $$
     where the parameters $\alpha \in \mathbb{R}_+$ and $\beta \in \mathbb{R}_{+}$ tune the strength of the 
     global negative feedback and
     local positive feedback provided by the current source,
     respectively. 
     Since $f = k_v \, V$ with $k_v >0$, positive (negative) voltage leads to crawler extension (contraction).\\
     \textit{(2) A proprioceptive feedback current} ($i_p$), inversely proportional to crawler's strain ($s$): 
     $$
         i_p = -\gamma (u_2 - u_1 ) = -\gamma \, s,
         \label{eq:i_p}
     $$
     where $\gamma \in \mathbb{R}_+$ is the proprioceptive feedback gain in the voltage dynamics. $i_p$ makes the output voltage grow (decrease) as the crawler contracts (extends). Thus, the resulting closed-loop electromechanical interaction is of negative feedback kind.

For convenience, we introduce a change of coordinates\footnote{
In coordinates
$[V \; \dot{u}_{\text{com}} \; s \; \dot{s}]^{\top}$, 
all entries are  periodic during crawling (i.e., we leverage the translation invariance
to avoid working with absolute displacements).
} and write the dynamics \eqref{eq:dimensional_crawler_dynamics} in the strain $s:=u_2 - u_1$  of the crawler and the displacement of its center of mass $u_{\text{com}}:= (u_1 + u_2)/2$.  
The resulting closed loop is
\begin{subequations}
\begin{align}
    c \, \dot V  &=  -\alpha \,  V^3 + \beta \, V - \gamma \, s + i, \label{eq:dimensional_voltage_dynamics}\\
   m  \ddot{u}_{\text{com}} & =  - \frac{1}{2} \Big[ f_f\Big(\dot{u}_{\text{com}} - \frac{\dot{s}}{2}\Big) - f_f\Big( \dot{u}_{\text{com}} + \frac{\dot{s}}{2} \Big) \Big], \label{eq:dimensional_ucom_dynamics}\\
   m\ddot{s} & =  f_f\Big( \dot{u}_{\text{com}} - \frac{\dot{s}}{2} \Big) - f_f\Big(\dot{u}_{\text{com}} + \frac{\dot{s}}{2}\Big) \nonumber \\
   & \hspace{3.7cm} -2 (k s + b \dot{s}  - k_v V), \label{eq:dimensional_s_dynamics}
\end{align}
\label{eq:dimensional_electromechanical_dynamics}
\end{subequations}
where $c$ is a capacitance and $i$ an external applied current, as shown in Fig. \ref{fig:crawler_schematics}b). $i$ can be used, for example, to initiate crawling from equilibrium; however, it does not play a role in \eqref{eq:dimensional_electromechanical_dynamics} being able to self-sustain oscillations. Thus, for simplicity, we set $i \equiv 0$ for the remaining of this work. 

\vspace{-.25cm}
\subsection{Nondimensionalization}
We nondimensionalize the closed-loop electromechanical dynamics \eqref{eq:dimensional_electromechanical_dynamics}. Nondimensionalization leads to a reduction in the dimension of the parameter space, 
to establish dominant terms in the dynamics,
and to the definition of \textit{dimensionless groups} with physical interpretation. 
By Buckingham's $\Pi$-theorem, the dimensionless counterpart of \eqref{eq:dimensional_electromechanical_dynamics} has \textit{8 dimensionless groups}.

We define the \textit{characteristic mechanical scales} in the system as 
\begin{equation}
   l_* = l_0,  \; m_* = 2\, m, \text{ and } t_* = \sqrt{\frac{m}{2 \, k}}. 
\end{equation}
 $t_*$ is the inverse of the crawler's strain (undamped) natural frequency. We use 
 $V_*$ to denote the characteristic scale of the voltage; finding it requires some further considerations and we set it later. According to the characteristic scales, we define \textit{dimensionless variables}\footnote{We use $\mathsf{sans \; serif}$ style
 to denote dimensionless variables.  Dimensionless groups are denoted by $\pi_{\nd{i}}$, except for the damping ratio
 $\zeta$. } 
\begin{equation}
    \nd{t} := t / t_* , \; \nd{s} := s/l_*, \text{ and } \nd{V} := V / V_*.
\label{eq:dimensionless_variables}
\end{equation} 
We use $\nd{v_{\text{com}}}$ to denote the speed of the center of mass and $\nd{v_s}$ to denote the strain rate. We define the state $\boldsymbol{\mathsf{x}}(\mathsf{t})$ as $\boldsymbol{\nd{x}}(\nd{t}) = [\nd{V}\nd{(t)} \; \nd{v}_{\text{com}}\nd{(t)} \; \nd{s}\nd{(t)} \; \nd{v_s}\nd{(t)}]^{\top}$.
Accordingly, the dimensionless counterpart of the closed loop \eqref{eq:dimensional_electromechanical_dynamics} is
\begin{subequations}
    \begin{align}
        \nd{V}^{\prime}  &= - \pi_{\nd{c}} \,  \nd{V}^3 +\pi_{\nd{l}} \, \nd{V} - \pi_{\nd{s}} \, \nd{s}, \label{eq:V_dynamics}\\
  \nd{v}_{\text{com}}^{\prime} &=  - \frac{\pi_{\nd{f}}}{2} \, \Big( \sigma_{\pi}\Big(\nd{v}_{\text{com}} - \frac{\nd{v_s}}{2}\Big) + \sigma_{\pi}\Big(\nd{v}_{\text{com}} + \frac{\nd{v_s}}{2}\Big)  \Big), \label{eq:CoM_dynamics}\\
  \nd{s}^{\prime} & = \nd{v_s}, \\
   \nd{v_s}^{\prime} &=  \pi_{\nd{f}} \Big( \sigma_{\pi} \Big(\nd{v}_{\text{com}} - \frac{\nd{v_s}}{2}\Big) - \sigma_{\pi} \Big(\nd{v}_{\text{com}} + \frac{\nd{v_s}}{2}\Big) \Big)  \nonumber \\
   & \hspace{4cm} - \nd{s}  - 2 \, \zeta \, \nd{v_s}   + 2 \pi_{\nd{v}} \nd{V}, \label{eq:S_dynamics}
\end{align}
\label{eq:dimensionless_electromechanical_dynamics_movingFrame}
\end{subequations}
where $\sigma_{\pi}$ is as \eqref{eq:sigma_def}, but with dimensionless parameter $\pi_{\epsilon}$ rather than $\epsilon_f$, and $(\cdot)^{\prime}$ denotes $\mathrm{d}(\cdot)/\mathrm{d}\nd{t}$. The definitions of the dimensionless groups in \eqref{eq:dimensionless_electromechanical_dynamics_movingFrame} are provided in Table \ref{table:system_params}. $\zeta$ is the damping ratio of the crawler; $\pi_{\nd{f}}$ and $\pi_{\nd{v}}$ are the ratios of the frictional and feedback actuator forces to the elastic force, respectively. $\pi_{\epsilon}$ and $n_f$ are the slope and anisotropy parameters of the dimensionless friction model. $\pi_{\nd{c}}$, $\pi_{\nd{l}}$, and $\pi_{\nd{s}}$ in the voltage dynamics \eqref{eq:V_dynamics} can be interpreted as ratios of mechanical to electric timescales.

We note that the closed-loop dynamics \eqref{eq:dimensionless_electromechanical_dynamics_movingFrame} is $\Phi$-equivariant, where $\Phi$ denotes the linear mapping 
\begin{equation}
    \Phi\big([\nd{V} \; \nd{v_{com}} \; \nd{s} \; \nd{v_s}]^{\top}\big) := [-\nd{V} \; \nd{v_{com}} \; -\nd{s} \; -\nd{v_s}]^{\top}.
    \label{eq:symmetry}
\end{equation}
Any self-sustained periodic orbit in the system will inherit the same symmetry.

\begin{table}[h!]
\begin{center}
\caption{ \begin{small} (Left) Dimensional parameters of the electromechanical closed-loop crawler and their base units: mass ($M$), length ($L$), time ($T$), and current ($I$). (Right) Definition of the dimensionless groups in \eqref{eq:dimensionless_electromechanical_dynamics_movingFrame}.
 \end{small}}
 \label{table:system_params}
\begin{tabular}{|c | c|} 
 \hline
 Dimensional & Base    \\
 parameter & Units  \\ [0.5ex] 
 \hline\hline
 $m$ & $M$  \\ 
 \hline
 $l_0$ & $L$  \\
 \hline
 $k$ &  $M T^{-2}$   \\
 \hline
 $b$ &  $M T^{-1}$ \\
 \hline
 $f_{max}$ & $M L T^{-2}$  \\
  \hline
$k_v$ &  $T I L^{-1}$  \\ 
 \hline
$c$ &  $M^{-1} L^{-2} T^4 I^2$ \\ 
 \hline
$\alpha$ & $M^{-3} L^{-6} I^4 T^9$ \\ 
 \hline
 $\beta$ & $I^2 T^3 M^{-1} L^{-2}$ \\ 
  \hline
$\gamma$ & $I L^{-1}$ \\ 
 \hline
 $\epsilon_f$ & $L T^{-1}$ \\
 \hline
\end{tabular}
\quad
\begin{tabular}{|c|}
\hline 
Dimensionless\\
Group \\
\hline \hline
 $ \zeta  := b/\sqrt{2mk}$ \\
 \hline
 $\pi_{\nd{f}}  := f_{max}/( 2\, k \, l_*)$ \\
 \hline
 $\pi_{\nd{v}}  := \frac{1}{2} \,  k_v V_* / (k \, l_*)$ \\
 \hline
 $\pi_{\epsilon} := l_*/(t_* \epsilon_f) $ \\
 \hline
     $n_f$ \\
    \hline
 $\pi_{\nd{c}} := \alpha \, V_*^2 t_* / c  $ \\
 \hline
 $\pi_{\nd{l}} := \beta t_*/ c $ \\
 \hline
  $\pi_{\nd{s}}  := \gamma \, l_*  t_* / (c \, V_* )$ \\
  \hline
\end{tabular}
\end{center}
\end{table}


\subsection{Timescale separation: fast \& slow dynamics}

\begin{figure}[h!]
    \centering
    \includegraphics[width=8.4cm]{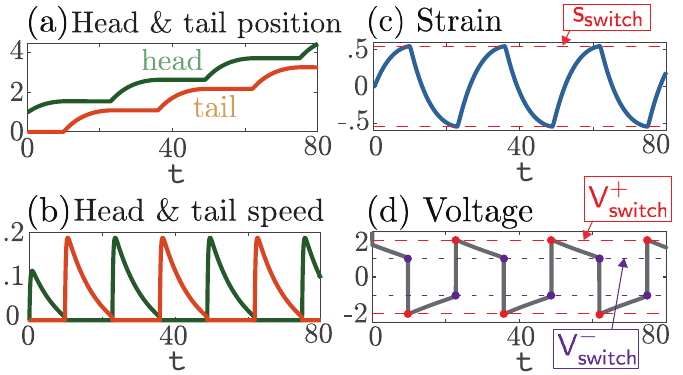}
    \caption{Trajectories of \eqref{eq:dimensionless_electromechanical_dynamics_movingFrame} obtained in numerical simulation with parameter values: $\zeta = 4.7, \pi_{\nd{f}} =  2.5, \pi_{\nd{v}} = 0.5 , \pi_{\nd{\epsilon}} = 4.7 \cdot 10^3, n_f = 1.5, \pi_{\nd{c}} = 10^4, \pi_{\nd{l}} = \pi_{\nd{s}}  = 2 \cdot 10^4$, and initial condition $\nd{V}(0) = 2, \nd{s}(0) = \nd{v_s}(0) = \nd{v_{com}}(0) = 0$. (a) and (b)  show trajectories and speeds of crawler's head and tail, respectively. (c) Strain levels in the crawler.
    (d) Relaxation oscillations in the voltage. The fast (slow) dynamics correspond to segments going from purple (red) to red (purple) dots. The values of $\nd{V^{+/-}_{switch}}$ are as defined in \eqref{eq:switching condition_voltage-} and \eqref{eq:switching condition_voltage+}. All plots display dimensionless variables.} 
    \label{fig:crawling_summary}
\end{figure}

We expect $V_* \sim 10^{-1}$ volts, $l_0 \sim 10^{-1}$m, $c \sim 10^{-6}$ farads, and $t_* \sim 10^{-1}$s. 
We set $\alpha \sim 10$ and $\beta, \gamma \sim 10^{-1}$, all in SI units, to yield reasonable strain levels in the crawler. Then, $\pi_{\nd{l}}, \pi_{\nd{c}}, \pi_{\nd{s}} \sim 10^4$. Consequently,  we define $V_* := 10^2 \sqrt{c/(t_* \, \alpha)}$. Since entries in the dimensionless state have been normalized to be of order 1 and dimensionless groups in the crawler dynamics are also expected to be of order 1, the large values of $\pi_{\nd{l}}, \pi_{\nd{c}}, \pi_{\nd{s}}$ imply that the voltage dynamics \eqref{eq:V_dynamics} are orders of magnitude faster than the crawler's. That is, there is \textit{a separation of timescales between electrical (controller) and crawler dynamics}. Such a separation of timescales becomes apparent in the \textit{relaxation oscillations} the output voltage experiences, shown in Fig. \ref{fig:crawling_summary}d). This kind of oscillations 
reflects the fast localized positive feedback (provided by the localized negative conductance current source $i_{\rm FHN}$) plus slow negative feedback (provided by the proprioceptive electromechanical feedback) nature of the designed crawler.

The timescale separation can be made explicit in the dynamics as follows:

\noindent \underline{\textit{Slow dynamics.}} $\nd{v}_{\text{com}}(\mathsf{t}), \nd{s}(\mathsf{t}), \nd{v_s}(\mathsf{t})$ are the \textit{slow state variables} and $\nd{t}$ is the \textit{slow timescale}.
\begin{subequations}
    \begin{align}
        \varepsilon \, \nd{V}^{\prime}  &= - \pi_{\nd{c}}^{(\varepsilon)} \,  \nd{V}^3 +\pi_{\nd{l}}^{(\varepsilon )} \, \nd{V} - \pi_{\nd{s}}^{(\varepsilon )} \, \nd{s}, \\
  \nd{v}_{\text{com}}^{\prime} &=  - \frac{\pi_{\nd{f}}}{2} \, \Big( \sigma_{\pi}\big(\nd{v}_{\text{com}} - \frac{\nd{v_s}}{2}\big) + \sigma_{\pi}\big(\nd{v}_{\text{com}} + \frac{\nd{v_s}}{2}\big)  \Big), \\
  \nd{s}^{\prime} & = \nd{v_s}, \\
   \nd{v_s}^{\prime} &=  \pi_{\nd{f}} \Big( \sigma_{\pi} \big(\nd{v}_{\text{com}} - \frac{\nd{v_s}}{2}\big) - \sigma_{\pi} \big(\nd{v}_{\text{com}} + \frac{\nd{v_s}}{2}\big) \Big)  \nonumber \\
   & \hspace{3.6cm} - \nd{s}  - 2 \, \zeta \, \nd{v_s}   + 2 \pi_{\nd{v}} \nd{V},
\end{align}
\label{eq:slowDynamics}
\end{subequations}
where $\varepsilon = 10^{-4}$ and $\pi_{\nd{i}}^{(\varepsilon)} := \varepsilon \, \pi_{\nd{i}}$ (with $\nd{i}= \nd{l}, \nd{c}, \nd{s})$ --  which are dimensionless groups of order 1.

\noindent \textit{\underline{Fast dynamics}.} We define the \textit{fast timescale} $\nd{T}$ as
$
   \nd{T} := \nd{t}/\varepsilon. 
$
$\nd{V}(\mathsf{T})$ is the \textit{fast state variable}. 
A change in timescale yields
\begin{subequations}
    \begin{align}
        \dot{\nd{V}} &= - \pi_{\nd{c}}^{(\varepsilon)} \,  \nd{V}^3 +\pi_{\nd{l}}^{(\varepsilon )} \, \nd{V} - \pi_{\nd{s}}^{(\varepsilon )} \, \nd{s}, \\
  \dot{\nd{v}}_{\text{com}} &=  - \varepsilon \frac{\pi_{\nd{f}}}{2} \, \Big( \sigma_{\pi} \big( \nd{v}_{\text{com}}- \frac{\nd{v_s}}{2}\big)  + \sigma_{\pi}\big( \nd{v}_{\text{com}} + \frac{\nd{v_s}}{2}\big) \Big), \\
  \dot{\nd{s}} & = \varepsilon \, \nd{v_s}, \\
   \dot{\nd{v}}_{\nd{s}} &=  \varepsilon \bigg[\pi_{\nd{f}} \Big( \sigma_{\pi} \big(\nd{v}_{\text{com}} - \frac{\nd{v_s}}{2}\big) - \sigma_{\pi} \big(\nd{v}_{\text{com}} + \frac{\nd{v_s}}{2}\big) \Big)  \nonumber \\
   & \hspace{3.6cm} - \nd{s}  - 2 \, \zeta \,  \nd{v_s}   + 2 \pi_{\nd{v}} \nd{V}\bigg], 
\end{align}
\label{eq:fastDynamics}
\end{subequations}
where abusing notation we denoted $\mathrm{d} \nd{x}/\mathrm{d} \nd{T}$ by $ \dot{\nd{x}}$. 

The slow \eqref{eq:slowDynamics} and fast \eqref{eq:fastDynamics} dynamics are amenable to \textit{geometric singular perturbation analysis}, which provides further insights into the geometry of the orbits of the closed-loop dynamics -- see \cite{Jones1995}. Letting $\varepsilon \to 0$:

\noindent \textit{\underline{Singularly perturbed slow dynamics}.} $\nd{v}_{\text{com}}, \nd{s}$, and $\nd{v_s}$ evolve over the \textit{slow manifold} defined by \eqref{eq:slow_manifold}: 
\begin{subequations}
    \begin{align}
        0  &= - \pi_{\nd{c}}^{(\varepsilon)} \,  \nd{V}^3 +\pi_{\nd{l}}^{(\varepsilon )} \, \nd{V} - \pi_{\nd{s}}^{(\varepsilon )} \, \nd{s}, \label{eq:slow_manifold}\\
  \nd{v}_{\text{com}}^{\prime} &=  - \frac{\pi_{\nd{f}}}{2} \, \Big( \sigma_{\pi}\big(\nd{v}_{\text{com}} - \frac{\nd{v_s}}{2}\big) + \sigma_{\pi}\big(\nd{v}_{\text{com}} + \frac{\nd{v_s}}{2}\big)  \Big), \\
  \nd{s}^{\prime} & = \nd{v_s}, \\
   \nd{v_s}^{\prime} &=  \pi_{\nd{f}} \Big( \sigma_{\pi} \big(\nd{v}_{\text{com}} - \frac{\nd{v_s}}{2}\big) - \sigma_{\pi} \big(\nd{v}_{\text{com}} + \frac{\nd{v_s}}{2}\big) \Big)  \nonumber \\
   & \hspace{3.6cm} - \nd{s}  - 2 \, \zeta \, \nd{v_s}   + 2 \pi_{\nd{v}} \nd{V},
\end{align}
\label{eq:slowDynamics_singularPerturbation}
\end{subequations}
\noindent \textit{\underline{Singularly perturbed fast dynamics}.}
$\nd{V}$ evolves while $\nd{v_{\text{com}}}, \nd{s}$, and $\nd{v_s}$ remain constant:
\begin{subequations}
    \begin{align}
        \dot{\nd{V}} &= - \pi_{\nd{c}}^{(\varepsilon)} \,  \nd{V}^3 +\pi_{\nd{l}}^{(\varepsilon )} \, \nd{V} - \pi_{\nd{s}}^{(\varepsilon )} \, \nd{s},\\
  \dot{\nd{v}}_{\text{com}} &=  0, \;\; \dot{\nd{s}}  = 0, \;\;\;  \dot{\nd{v_s}} = 0.
\end{align}
\label{eq:fastDynamics_singularlyPerturbed}
\end{subequations}

\subsection{Emergence of self-sustained oscillations}
Analysis of the singularly perturbed dynamics sheds light on the
mechanism that leads to a limit cycle in the closed-loop system, and the key role that the bistability of the voltage dynamics plays in generating it. 

\begin{figure}
    \centering
    \includegraphics[width=8.4cm]{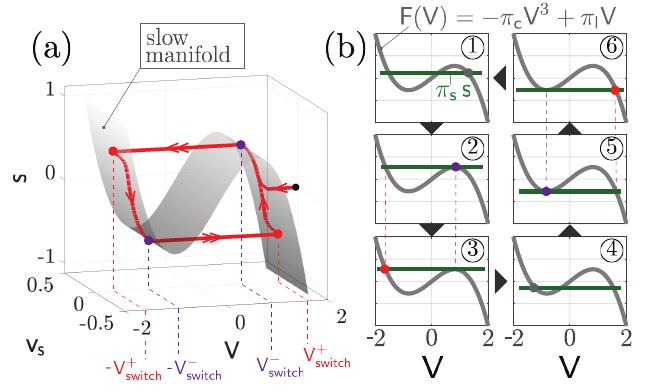}
    \caption{
    (a) Limit cycle in the state variables $\nd{s}, \nd{v_s}, \nd{V}$ in red. Single (double) arrows on the limit cycle correspond to trajectories of the slow (fast) dynamics.  The slow manifold \eqref{eq:slow_manifold} is represented in grey. The switching points are indicated in purple; the initial condition is in black. (b) Schematic of the strain-modulated switching. Circled numbers indicate the temporal ordering of the snapshots. The value of the voltage is indicated by the dot in each panel. Purple (red) dots correspond to the value of the voltage right before (after) the switch.
    Parameter values and initial condition are as in Fig. \ref{fig:crawling_summary}. } 
    \label{fig:limit_cycle}
\end{figure}

In the slow dynamics \eqref{eq:slowDynamics_singularPerturbation}, the state evolves along the manifold \eqref{eq:slow_manifold}, where $\mathsf{V}$ lies at a \textit{stable} fixed point -- see Fig. \ref{fig:limit_cycle}a). This fixed point is modulated by the strain $\mathsf{s}$ in the crawler. 
When the strain reaches a critical thresholding value 
    such that $\pi_{\mathsf{s}}^{(\varepsilon)}  \mathsf{s}$ is tangent to $\mathsf{F(V)} = -\pi_{\mathsf{c}}^{(\varepsilon)} \mathsf{V}^3 + \pi_{\mathsf{l}}^{(\varepsilon)} \mathsf{V}$,
    the fixed point becomes \textit{unstable}.  We refer to this criticality as the \textit{switching condition}, since $\mathsf{V}$ switches from one branch of the bistable manifold to the other, and the dynamics switch from slow \eqref{eq:slowDynamics_singularPerturbation} to fast \eqref{eq:fastDynamics_singularlyPerturbed}. The switching condition is characterized by
    \begin{equation}
    \nd{V^-_{switch}} = \pm \sqrt{\frac{\pi_{\mathsf{l}}}{3\, \pi_{\mathsf{c}}}}, 
    \label{eq:switching condition_voltage-}
    \end{equation}
    or equivalently, in terms of strain by
    $
    \nd{s_{switch}} = \pm \frac{2 \pi_{\mathsf{l}}^{3/2} }{3 \sqrt{3} \pi_{\mathsf{c}}^{1/2} \pi_{\mathsf{s}} } .
        \label{eq:switching condition_strain}
    $
   Once the switching condition is met, $\nd{V}$ evolves following \eqref{eq:fastDynamics_singularlyPerturbed}, yielding a spike -- see Fig. \ref{fig:crawling_summary}d) -- until it reaches the only \textit{stable} fixed point left in the system, located at the other branch of the slow manifold -- see Fig. \ref{fig:limit_cycle}. At this fixed point the value of $\nd{V}$ is
    \begin{equation}
         \nd{V^+_{switch}} = - 2 \, \nd{V^-_{switch}}.
    \label{eq:switching condition_voltage+}
    \end{equation}

    Once in the slow manifold, the state evolves again according to the slow dynamics \eqref{eq:slowDynamics_singularPerturbation}, until the switching condition is met -- this time with the opposite sign due to the symmetry $\Phi$ defined in \eqref{eq:symmetry}  -- activating the fast dynamics and repeating the cycle. 
    
    The strain-modulated \textit{bistable} electrical dynamics is the mechanism that makes the closed-loop system \eqref{eq:dimensionless_electromechanical_dynamics_movingFrame} generate periodic trains of switched voltage leading to peristaltic waves in the crawler. Since the orbits that $\nd{V}$ follows to travel between the two branches of the slow manifold are not the same, $\nd{V}$ behaves as an \textit{hysteretic switch}, as shown in Fig. \ref{fig:limit_cycle}a).

\section{Conclusion}
We have proposed an \textit{excitable nonlinear controller} -- inspired by the FitHugh-Nagumo model of neural dynamics -- to control the peristaltic locomotion of a soft one-segmented robotic crawler. Dimensional analysis and nondimensionalization of the closed-loop dynamics reveal a \textit{separation of timescales} in the closed-loop system, which leads to \textit{relaxation oscillations} in the voltage. We leveraged the separation of timescales for singular perturbation analysis of the closed-loop dynamics. Combining the geometric insights that singular perturbation analysis provides and numerical simulations, we shed light on the mechanism that leads to self-sustained peristaltic motion in the closed-loop crawler, illustrating the feasibility of spiking control to engineer the locomotion of soft-bodied robots.

Ongoing work includes the characterization of bifurcations in the closed-loop system and the design of an online adaptation strategy inspired  by neuromodulation
-- \cite{Schmetterling2022,Ribar2019,Marder2012,Marder2014} --
for the electrical scales to tune the periodic crawling motion as environmental or other conditions change. It is expected that by matching scales from the electrical dynamics to the mechanical dynamics, increased efficiency of the closed-loop system will be achieved. Further efficiency and environmental awareness in excitable crawling could be accomplished by applying the external current $i$, which in this work was taken to be $i \equiv 0$ for simplicity. This current could be utilized, for example, to incorporate \textit{extrapropioceptive feedback from other sensor modalities} carrying environmental information useful for decision-making and crawling gait modulation. Future work includes analyzing a distributed multi-segmented crawler, where peristaltic waves will be sustained using only \textit{local} proprioceptive information exchange. We will leverage the inter-segmental connectivity pattern of the crawler for this aim -- following ideas from \cite{Arbelaiz2020, Arbelaiz2021, Arbelaiz2022,Arbelaiz2023,Bamieh,Epperlein2016} and \cite{Leonard2024}.




\newpage
\bibliography{ifacconf}             
                                                   







\end{document}